\documentclass[aps,prd,amssymb,nofootinbib,floatfix, a4paper,twocolumn]{revtex4}

\usepackage{mathrsfs}
\usepackage{latexsym,amsmath,amsfonts,amssymb}
\usepackage{bm}

\usepackage{graphicx}

\allowdisplaybreaks

\newcommand{\be}{\begin{equation}}
\newcommand{\ee}{\end{equation}}
\newcommand{\beq}{\begin{equation}}
\newcommand{\eeq}{\end{equation}}
\newcommand{\bea}{\begin{eqnarray}}
\newcommand{\eea}{\end{eqnarray}}
\newcommand{\eq}{\eqref}

\begin{document}

\title{Editorial note to Jean-Marie Souriau's \\
`` On the motion of spinning particles in general relativity"}

\author{Thibault Damour}
\email{damour@ihes.fr}
\affiliation{Institut des Hautes Etudes Scientifiques, 35 route de Chartres, 91440 Bures-sur-Yvette, France}

\date{\today}

\begin{abstract} 
The gravitational interaction of (classical and quantum) spinning bodies is currently the focus of many works using
a variety of approaches.  
This note is a comment on a short paper by Jean-Marie Souriau, now reprinted in the GRG Golden Oldies collection.
Souriau's short 1970 note was a pioneering contribution to a symplectic
description of the dynamics of spinning particles in general relativity which remained somewhat unnoticed.
We explain the specificity of Souriau's approach and emphasize its potential interest within the current flurry of
activity on the gravitational interaction of spinning particles.
 \end{abstract}

\maketitle

This Golden Oldie is the English translation of Souriau's article \cite{SouriauCRAS1970}, which was 
originally published in French as a 3-page note to the Comptes Rendus of
the French Acad\'emie des Sciences . It is the first publication of Souriau on the general-relativistic 
generalisation of his special-relativistic (and Galilean-relativistic) symplectic approach to spinning particles, as presented  in his 1970 book on the structure of dynamical systems \cite{SouriauSSD}. Strangely enough
though Souriau's article was the first to present the pre-symplectic 2-form $\sigma$ describing the dynamics of spinning
particles coupled to an Einsteinian curved background, it remained largely unnoticed in the subsequent literature on
symplectic, or canonical, approaches to spinning particles. In particular, K\"unzle's important 1972 paper on the
``Canonical Dynamics of Spinning Particles in Gravitational and Electromagnetic Fields" \cite{Kuenzle:1972uk},
which was the first to define a pre-symplectic ``potential" for $\sigma$ (i.e. a Cartan 1-form, $\varpi$,  such that
$\sigma = - d \varpi$) does not cite Souriau's 1970 note.  It is also not cited in connection with the construction
of a symplectic  formalism  in Souriau's 1974 long
paper on spinning particles in electromagnetic and gravitational fields \cite{Souriau1974}.

Let us briefly discuss the main results presented in Souriau's note, with an emphasis on its most novel contribution, namely the
pre-symplectic 2-form $\sigma$ describing the dynamics of spinning particles in general relativity\footnote{We use a mostly plus signature, and usual conventions for the components (in a coordinate system $x^\mu$, $\mu=0,1,2,3$) of the curvature tensor of the metric tensor $g_{\mu \nu}$ (with associated Levi-Civita connection $\nabla$), namely:
 ${R^{\alpha}}_{\beta \mu \nu} \equiv \partial_\mu
  \Gamma^{\alpha}_{\beta \nu}- \partial_\nu  \Gamma^{\alpha}_{\beta \mu}+ \Gamma^{\alpha}_{  \sigma \mu} \Gamma^\sigma_{\beta \nu}
  -  \Gamma^{\alpha}_{  \sigma \nu} \Gamma^\sigma_{\beta \mu}$.}.

First, Souriau presents the skeletonized description of a pole-dipole particle in general relativity
by a first-order distribution,
\be \label{F1}
 {\cal F}[\gamma_{\mu \nu}(x)]=\int \sqrt{- g} d^4 x T^{\mu \nu}(x) \gamma_{\mu \nu}(x),
 \ee
where $\gamma_{\mu \nu}(x)$ is a symmetric tensor-valued test function
 having its support on a curve, $\Gamma$, namely 
\be \label{F2}
{\cal F}[\gamma_{\mu \nu}(x)]=\int_{\Gamma} \left[P^\mu \gamma_{\mu \nu} + S^{\lambda \mu} \nabla_\lambda \gamma_{\mu \nu} \right] dX^\nu\,.
\ee
He then derives the associated universal Mathisson-Papapetrou equations \cite{Mathisson:1937zz,Papapetrou:1951pa}, namely
(with $U^\mu \equiv d X^\mu(s)/ds$, $s$ being an arbitrary parameter along the curve)
\begin{eqnarray} \label{MP}
     \frac{ \nabla S^{\mu \nu }}{ds} &=&P^\mu U^\nu -P^\nu U^\mu\,,\\
   \frac{\nabla P^\mu}{ds} &=&- \frac12 {R^\mu}_{\nu \alpha \beta} U^\nu S^{\alpha \beta} \,,
\end{eqnarray}
from the condition that the functional ${\cal F}[\gamma_{\mu \nu}(x)]$ vanishes when $\gamma_{\mu \nu}$
is the Lie derivative of $g_{\mu \nu}$ (i.e.
$\gamma_{\mu \nu}= \nabla_\mu V_\nu+ \nabla_\nu V_\mu$), with a compact-support vector field $V^\mu$.

Let us note that this method of derivation of the dynamics of  a pole-dipole particle in general relativity is (essentially)
equivalent to the one used by Mathisson in 1937 \cite{Mathisson:1937zz}. The distributional aspect of this derivation
was made more explicit by Tulczyjew in 1959  \cite{Tulczyjew59} who wrote down the distributional stress-energy tensor 
corresponding to Eqs. \eq{F1}, \eq{F2}, namely
\bea
 T^{\mu \nu}(x) &=& \int_{\Gamma} ds P^{(\mu} U^{\nu)} \hat\delta(x-X(s))  \nonumber\\
&& - \nabla_\lambda \left[ \int_{\Gamma} ds S^{\lambda (\mu} U^{\nu)} \hat\delta(x-X(s))  \right]\,,
\eea
where $A^{(\mu \nu)}\equiv \frac12\left( A^{\mu \nu}+A^{\nu \mu}\right)$,
and where $\hat\delta(x-X) = \delta^{(4)}(x-X)/\sqrt{- g}$ is a (scalar-valued) 4-dimensional Dirac delta.

Mathisson's classic results have been reprinted (and translated) as Golden Oldies \cite{Mathisson1937GO}. An
 authoritarive review of Mathisson's programme and its later developments have been given by Dixon \cite{Dixon2008}.
We will therefore shun any further discussion of this aspect of Souriau's paper, except for mentioning the elegant
use of the functional \eq{F2} for deriving the value, for a spinning particle, of the general conserved quantity 
$\mu(K)= \int_\Sigma \sqrt{- g} d^4 x T^{\mu \nu}(x) K_\mu d\Sigma_\nu$ (integrated on an hypersurface $\Sigma$) 
 associated to any Killing vector $K$. It is obtained 
 by evaluating ${\cal F}$ on a test tensor of the form $\gamma_{\mu \nu}=K_{(\mu} \nabla_{\nu)} u$, with an arbitrary
 function $u(x)$ interpolating between 0 in the past and 1 in the future. This yields
 \be \label{mu}
 \mu(K)= P^\mu K_\mu + \frac12 S^{\mu \nu}  \nabla_{\mu} K_\nu\,.
 \ee
Though this derivation of the conserved quantity $\mu(K)$ is not based  on the symplectic version of
Noether's theorem, Souriau  uses here the notation $\mu$ to link it to the corresponding 
symplectic ``moment" of the Poincar\'e group discussed in his book \cite{SouriauSSD} (corresponding to 
a generic special-relativistic Killing vector $K_\mu(x) = a_\mu + b_{[\mu \nu]} x^\nu$).

The most important result of Souriau's paper is to give a {\it symplectic formulation}
of the Mathisson-Papapetrou evolution equations \eq{MP},
completed by the (Tulczyjew) spin-supplementary condition 
\be \label{SP=0}
 S^{\mu \nu} P_\nu=0\,.
\ee
In his book \cite{SouriauSSD}, Souriau had given a symplectic formulation of spinning particles in special relativity
(and Galilean relativity). [Souriau's formulation was originally inspired by a work of  Bacry \cite{Bacry1967}
 aimed at giving a group-theoretical
definition of an {\it elementary} classical spinning particle.]
 The basic ingredients of such a formulation are: (1) to represent the considered dynamics
as taking place in some {\it evolution space} ${\cal Y}$ (corresponding to the {\it extended phase space} $(y^A)=(t, q^i, p_i)$
of usual canonical formulations); and (2) to describe the equations of motion as the kernel of a closed 2-form 
$\sigma = \frac12 \sigma_{AB}(y) dy^A \wedge dy^B$
on the evolution space  ${\cal Y}$, i.e. as the curves $y^A(s)$ (where $s$ denotes an arbitrary parameter) such that
\be \label{sigmady=0}
 \sigma_{AB}(y) \frac{d y^B}{ds }=0\,.
\ee
The 2-form $\sigma $ should be pre-symplectic, i.e it should be closed ($d\sigma =0$), and it should be of constant rank.
In the usual case of a $(2n+1)$-dimensional evolution space $(t, q^i, p_i)$ for $n$ degrees of freedom 
$q^i$ (with conjugate momenta $p_i$), the rank should be $2 n$, i.e. equal to the dimension of the {\it space of 
motions} (i.e. the set of solutions of the equations of motion, parametrized, say, by the values of $q^i$ and $p_i$ 
at some initial instant). A spinning particle should have 4 degrees of freedom,
indeed its ``motions" should be parametrized by 8 real numbers: 3 initial positions, 3 initial
velocities, and 2 angles for the initial direction of the spin vector. 
The pull back of $\sigma $ on the space of motions should then 
define a symplectic structure, i.e. be closed and non-degenerate. 

This symplectic formulation of Hamiltonian dynamics was pioneered by Cartan \cite{Cartan1922} who emphasized the important roles
of the (``Cartan") 1-form
\be \label{cartan}
\varpi  \; (= p_i dq^i- H(q,p,t) dt)\,,
\ee
and of its differential
\be
\sigma = -d \varpi  \; (= dq^i \wedge dp_i + dH\wedge dt)\,.
\ee
Here we put in parentheses the expressions of $\varpi$ and $\sigma$ in the usual canonical formalism, based on the use of
Darboux coordinates $(q^i, p_i)$ [and we used the convention where the symplectic form reads $ dq^i \wedge dp_i + \cdots$, corresponding to a Poisson bracket $\{q^i, p_j  \}= \delta^i_j$]. Cartan emphasized that these forms define, respectively, relative and absolute integral invariants
of the dynamics, considered in the evolution space (including time). He also showed that the equations of motion take the form \eq{sigmady=0}.
It is also important to keep in mind that Eq. \eq{cartan} shows that the Cartan 1-form $\varpi $, besides being a ``potential" from
which $\sigma$ is derived, is also the integrand of the {\it Hamiltonian action} in evolution space: 
\be \label{SH}
S_H[y] = \int_{y_1}^{y_2} \varpi   (= \int [p_i dq^i- H(q,p,t) dt] )\,.
\ee
Souriau's symplectic formulation of spinning particles  is quite economical compared to most formalisms used in the
literature on the dynamics of spinning particles in special, or general, relativity, notably because: (i) it uses a minimal evolution space,
having 9 dimensions, leading, after quotienting the one-dimensional leaves solving Eq. \eq{sigmady=0}, to a correspondingly minimal
space of motions having 8 dimensions; and (ii) it directly defines a symplectic structure on the 8-dimensional space of motions.
A general point in Souriau's 9-dimensional evolution space, say ${\cal Y}_9$, is
\bea \label{Y9}
{\cal Y}_9: \;\; y&=&(X^\mu, I^\mu, J^\mu) \; \; {\rm with} \nonumber \\
 g_{\mu \nu}  I^\mu  I^\nu& =&-1\, ,\,g_{\mu \nu}  J^\mu  J^\nu=+1\, , \, g_{\mu \nu}  I^\mu  J^\nu=0.\nonumber\\
\eea
The unit timelike vector $ I^\mu$ describes the direction of the 4-momentum $P_\mu = m I_\mu$ , while
the unit spacelike vector $J^\mu$ (orthogonal to $I^\mu$)  denotes the direction of the spin 4-vector $S^\mu = s J^\mu$. 
Here, $m$ (the mass) and $s$ (the spin) are two constants. Souriau's presymplectic 2-form\footnote{Note that Souriau likes to write
a 2-form, say $dq^i \wedge dp_i$, as an explicit antisymmetric bilinear form of two independent infinitesimal variations, $dy=(dq^i, dp_i)$,
$\delta y=(\delta q^i, \delta p_i)$, in $(q^i, p_i)$ space, namely: $dq^i \delta p_i- \delta q^i dp_i$. Such a notation (inherited from
the classic D'Alembert idea of virtual displacements) was also used by Cartan \cite{Cartan1922}.}
 is then given by
\bea \label{sigma9}
&&\sigma= dX^\mu \wedge \nabla P_\mu + \frac1{2 s^2} S^{\mu \nu} g^{\alpha \beta}\nabla S_{\mu \alpha} \wedge \nabla S_{\nu \beta}
\nonumber\\
&&+ \frac14 S^{\mu \nu} R_{\mu \nu \alpha \beta} dX^\alpha \wedge dX^\beta\,.
\eea
Here, and below, the notation $\nabla$ denotes a covariant version of Cartan's exterior differential $d$ when acting on the
components of a tensor (each component being viewed as a 0-form in the evolution space ${\cal Y}$). E.g.,  $\nabla S_{\mu \nu}= d  S_{\mu \nu} - S_{\sigma \nu}\Gamma^\sigma_{\mu \lambda} dX^\lambda
- S_{\mu \sigma }\Gamma^\sigma_{\nu \lambda} dX^\lambda$ (it is not needed in the first term if one
uses a coordinate basis: $  dX^\mu \wedge \nabla P_\mu =  dX^\mu \wedge d P_\mu $). The spin tensor is defined as $S_{\mu \nu} \equiv s \eta_{\mu \nu \kappa \lambda} I^\kappa J^\lambda$,
where $\eta_{\mu \nu \kappa \lambda}$ denotes the Levi-Civita tensor with $\eta_{0123}= +\sqrt{-g} $.
As shown explicitly in \cite{Kuenzle:1972uk,Souriau1974} the kernel of $\sigma$ (i.e. the evolution equations 
\eqref{sigmady=0}) yield uniquely defined evolution equations on the 9-dimensional space \eqref{Y9} which are
equivalent to the MP equations \eqref{MP}, together with the following evolution equation for $X^\mu$
\be
m \frac{dX^\mu}{ds} =\left(\frac{- P_\nu \, dX^\nu}{m \, ds} \right) \left( P^\mu + \frac{1}{2 \Delta} S^{\mu\nu}R_{\nu\lambda \alpha \beta}P^\lambda S^{ \alpha \beta} \right)\,,
\ee
with a denominator equal to
\be
\Delta= m^2 +\frac14 R_{\mu \nu \alpha \beta} S^{\mu\nu} S^{ \alpha \beta}\,.
\ee

Souriau insists on stating that ``$\sigma$ is not derivable from a potential, thus {\it a fortiori} from a Lagrangian". Here, he means
that, though $\sigma$ is closed ($d \sigma=0$), and can therefore be {\it locally} derived from a Cartan 1-form ($ \sigma= -d \varpi$),
it is not {\it exact}, i.e. there does not exist  a {\it globally defined} 1-form $\varpi$ on  ${\cal Y}$ such that $ \sigma= -d \varpi$. 
The basic reason (given in \cite{SouriauSSD} and \cite{Souriau1974}) is that the restriction of $ \sigma$ to the 2-sphere of 
spin directions $(\theta, \phi)$ is equal to $ \sigma_s= s \sin \theta d \theta\wedge d\phi$. The integral of  $ \sigma_s$ over
all spin directions does not vanish (being equal to $4 \pi s$), therefore $ \sigma_s$ is not exact. This led Souriau to apply his
geometric quantization approach to spinning particles (see \cite{Souriau1974}), from which he deduced the constraint that
the spin value $s$ must be an integer multiple of $\frac{\hbar}{2}$. [Indeed, when 
 $ s = n \frac{\hbar}{2}$, for some integer $n$, the  $4 \pi s$ ambiguity in
the definition of the Hamiltonian action Eq. \eq{SH} becomes an unobservable multiple of $2 \pi$ in the
(quasi-classical) quantum phase $S_H/\hbar$.]

In spite of the distaste of Souriau for the use of (non globally defined)  potentials $\varpi$,  it is convenient to be able to describe the dynamics of classical spinning particles by means of the Hamiltonian action defined by some Cartan 1-form $\varpi$ (such that  $ \sigma= -d \varpi$).
This was first done by K\"unzle \cite{Kuenzle:1972uk}. In order to be able to work with a globally defined $\varpi$, Ref. \cite{Kuenzle:1972uk}
extended the minimal 9-dimensional evolution space ${\cal Y}_9$,  Eq. \eq{Y9}, used by Souriau into a 10-dimensional evolution space, say ${\cal Y}_{10}$, with elements
\bea \label{Y10}
{\cal Y}_{10}: \;\;  y=(X^\mu, I^\mu, J^\mu, K^\mu, L^\mu)\,,
 \eea 
keeping the identifications $P_\mu = m I_\mu$, and $S^\mu = s J^\mu$.
Here the two extra unit (and orthogonal) vectors  $ K^\mu, L^\mu$ complete the orthogonal dyad $ I^\mu, J^\mu$ into a full orthonormal
spacetime tetrad, say 
\be \label{tetrad}
(e^\mu_a)_{a=0,\ldots,3}= (I^\mu,  K^\mu, L^\mu, J^\mu).
\ee
  Note that this adds only one real parameter to the evolution space  ${\cal Y}$. Indeed, in the minimal 9-dimensional
evolution space  ${\cal Y}_9$, Eq. \eq{Y9}, the spin direction $ J^\mu$ needed (after having chosen $X^\mu$ and $I^\mu$)
two angles, say  $(\theta, \phi)$, for its specification (with
respect to the base manifold). The specification of the two extra unit vectors  $ K^\mu, L^\mu$ requires only the choice of an
extra angle $\psi$, measuring, say, the rotation of the two vectors  $ K^\mu, L^\mu$ in the 2-plane orthogonal to  $ I^\mu$ and $J^\mu$.
[When sitting in the ``rest frame" defined by $P_\mu = m I_\mu$, one only needs a third Euler angle, say $\psi$,
to specify a spatial rotation around the spin vector as axis of rotation.] When using as evolution space the 10-dimensional  manifold\footnote{${\cal Y}_{10}$ is the fiber
bundle of local Lorentz frames $(e^\mu_a)_{a=0,\ldots,3}$ over the spacetime manifold $X^\mu$, endowed with the metric $g_{\mu \nu}(X)$.} defined by  ${\cal Y}_{10}$, Eq. \eq{Y10},  K\"unzle \cite{Kuenzle:1972uk} could construct a globally defined potential $ \varpi_{10}$
for $ \sigma_{10}=- d \varpi_{10}$, namely
\be \label{cartan10}
\varpi_{10}= P_\mu  dX^\mu- s \omega_{KL} = m I_\mu  dX^\mu- s e_K  \cdot\nabla e_L\,,
\ee
where the second term, proportional to the spin magnitude $s$, is the $K, L$ 
component of the connection 1-form ${\omega^a}_b$ (in the frame bundle)
measuring the infinitesimal rotation (compared to the parallel transport
defined by  the metric $g_{\mu \nu}(X)$) of the tetrad \eq{tetrad}
as one moves from the fiber above $X^\mu$ to the fiber above $X^\mu + d X^\mu$,  namely
\be
\nabla e^\mu_a \equiv  d e^\mu_a + dX^\lambda \Gamma^\mu_{\nu \lambda}e^\nu_a  \equiv   e^\mu_b \, {\omega^b}_a \,.
\ee 
Here, $d e^\mu_a $ should not be thought of as something like $dX^\lambda \partial_\lambda  e^\mu_a (X)$
but as a 1-form in all the variables $X^\mu, I^\mu, \theta, \phi, \psi$ needed to label a generic (fiber bundle) element $y \in {\cal Y}_{10}$
(in the same way that $dp_i$ is independent from $dq^i$ and $dt$ in the usual extended phase space $(t,q^i,p_i)$).
We have added a subscript $10$ in Eq. \eq{cartan10} as a reminder that $\varpi_{10}$ and $ \sigma_{10}=- d \varpi_{10}$
are defined in the 10-dimensional evolution space ${\cal Y}_{10}$, Eq. \eq{Y10}. It is, however, easily seen that the 10th coordinate $\psi$
only enters $\varpi_{10}$ through the additional contribution $ s d\psi$. The work of   K\"unzle \cite{Kuenzle:1972uk} then
showed that the pull back on ${\cal Y}_{9}$ of  $ \sigma_{10}=- d \varpi_{10}$  on ${\cal Y}_{10}$ 
(under the natural projection of ${\cal Y}_{10}$ onto ${\cal Y}_{9}$) is equal to Souriau's 2-form \eq{sigma9}. The
corresponding equations of motion in ${\cal Y}_{10}$ are found to be equivalent to those in ${\cal Y}_{9}$, after ignoring
the fact that the evolution of the ignorable angle $\psi$ remains arbitrary. 

A useful consequence of the ignorability of $\psi$ is that, if one is ready to use an action for a spinning particle
having some non-uniqueness, but being defined directly on the minimal 9-dimensional evolution space ${\cal Y}_{9}$, one
can simply use the pull back of $\varpi_{10}$, Eq. \eq{cartan10}, under any (reasonably smooth) projection of ${\cal Y}_{10}$
onto the minimal evolution space ${\cal Y}_{9}$. From a practical point of view, this means for instance that if the 
spacetime manifold is endowed with a field of orthonormal frames (a  tetrad field or ``vierbein"), say $ E^\mu_a $ 
(with $a=0,1,2,3$), one can use this
tetrad field successively: (1) to describe the three independent components of the momentum direction vector $I^\mu$ (e.g. via
the triad components $P_i= m I_\mu  E^\mu_i $, where $i=1,2,3$); 
 (2) to specify the two independent components of the direction $J^\mu$ (orthogonal to $I^\mu$), e.g. via the two polar angles 
 $ \theta, \phi$ of the unit vector  ${\hat J}^\mu = B_I(J^\mu)$, where $B_I$ denotes the (unique) Lorentz boost transforming
 $I^\mu$ into the ``lab" unit time vector $ E^\mu_0 $; and (3) to define $K^\mu$ such that its lab-boosted version 
  ${\hat K}^\mu = B_I(K^\mu)$ is, say\footnote{We have in mind here using $ E^\mu_3 $ as z-axis for the polar coordinates.}, $ E^\mu_1 $.
  In this way, the nine coordinates $X^\mu, P_i, \theta, \phi$ suffice to specify a  point in ${\cal Y}_{10}$, and thereby to define
  a 9-dimensional pull back of $\varpi_{10}$, Eq. \eq{cartan10}, which provides an Hamiltonian action for the dynamics
  of a spinning particle in the curved spacetime endowed with the metric $g_{\mu \nu}(X)$.
  
 We spent time describing how the Souriau-K\"unzle approach directly defines, in the {\it minimal} 9-dimensional
 evolution space ${\cal Y}_{9}$ (where spin is described by {\it only two}  phase-space variables),
 both a pre-symplectic 2-form and an Hamiltonian action (or Cartan 1-form) to contrast it
 with  the other  formalisms  used in the literature, notably in  many recent papers (motivated
 by gravitational-wave physics) dealing with 
 gravitationally interacting systems of spinning black holes or neutron stars. 
  
  Indeed, many of these papers start with the use of non-minimal Lagrangian formalisms for spinning particles.
 Let us give some examples\footnote{Here, we will make no attempt at providing a thorough survey of the
 recent literature on  gravitationally interacting systems of spinning bodies. Our aim is only to provide a few entries
 in the literature to illustrate the relevance of Souriau's approach.}.  
 Refs. \cite{Porto:2005ac,Porto:2008tb,Barausse:2009aa} took as starting point 
the (special-relativistic) work of Hanson and Regge \cite{Hanson:1974qy}. This type of approach (see also 
Ref. \cite{Witzany:2018ahb}) uses
 a large (strongly nonminimal) initial evolution space having 20 dimensions: 4 positional coordinates $X^\mu$,
and their corresponding 4 velocities $\dot X^\mu =d X^\mu/ds$, or the momenta $P_\mu$, and 6 rotational coordinates $\phi_A$ (describing
a frame, or a Lorentz transformation $\Lambda^\mu_\nu(\phi_A)$ along the worldline) with their corresponding 6 angular velocities
${\sigma^\mu}_\nu= \dot \Lambda^\mu_\sigma \Lambda^\sigma_\nu$, or the
conjugate spin tensor $S_{\mu \nu}$.
This starting evolution space is endowed with initial Poisson brackets of the type
 $\{ x^\mu, P_\nu\} = \delta_\mu^\nu$, $\{\Lambda^\mu_\nu, S^{\alpha}_{ \beta}\}= \Lambda^\mu_\beta  \delta^\alpha_\nu- \Lambda^\mu_\beta  \delta^\alpha_\nu$, $ \{S^\mu_\nu, S^{\alpha}_{ \beta}\}= S^\mu_\beta  \delta^\alpha_\nu
+ 3 \; {\rm other \; terms}$.

 It is then necessary to impose many constraints (or to eliminate  the unphysical degrees of freedom in some other way)
to reduce the dimensionality of the evolution space down to the physical value 9 (instead of 20!).
This reduction is often done (following Ref. \cite{Hanson:1974qy}) by using Dirac's approach to constrained dynamics,
with its (rather heavy) technology of constraint-analysis  and corresponding definitions of Dirac brackets for the 
reduced degrees of freedom. It was also shown (e.g. \cite{Damour:2007nc,Hergt:2010pa,Schafer:2018kuf} and references therein) that the reduction of the spin degrees of freedom to a minimal canonical form
 could be performed within the Arnowitt-Deser-Misner framework. For methods of doing this reduction within the
 Effective-Field-Theory approach to spinning bodies see, e.g., 
Refs. \cite{Porto:2005ac,Levi:2015msa,Levi:2017kzq,Mandal:2022ufb,Levi:2022rrq}. The comparison with 
 the results to the Arnowitt-Deser-Misner approach have been  notably discussed in \cite{Levi:2014sba}.

Though there is nothing conceptually wrong in starting with a large initial evolution space, and in then
imposing constraints to reduce its dimensionality to the minimal needed one, it seems to the author of this 
editorial note that some of the complicated conceptual features in the various
reductions (and/or gauge-fixing) mentioned above would have benefitted from the use, from the start,
of the conceptually streamlined, minimal-dimension symplectic Souriau-K\"unzle approach\footnote{The recent 
approaches (e.g. \cite{Vines:2016unv,Vines:2017hyw}) use actions involving additional Lagrange multipliers contributions
that seem equivalent to imposing constraints in the strong Souriau-K\"unzle way.}. In addition, it seems that the symplectic approach is technically simpler than the constraint-plus-Dirac-bracket technology by its use of the
general property that $p$-forms, and their exterior differentials, have very simple (pull back) transformations
 under embedding transformations or restrictions to submanifolds. It has also the advantage of not fundamentally
 requiring the explicit use of canonically conjugate phase-space variables. Indeed, though Darboux's theorem
 guarantees that the symplectic form induced on the space of motions (or on
 any $X^0=$cst section of ${\cal Y}_{9}$) will always allow the (local) existence of canonical coordinates with respect to
 which the symplectic form takes the form\footnote{See, e.g., Section 4 of \cite{Bel1980} in the special-relativistic case.}
  $ dq^i \wedge dp_i + s \sin \theta d\theta \wedge d \phi= dq^i \wedge dp_i + s  d(-\cos \theta) \wedge d \phi$, one can also work with non-canonical variables, and simply use the
 general symplectic version of the equations of motion, of the type (when $\omega= \frac12\omega_{AB}(x) dx^A\wedge dx^B+ dH(x)\wedge dt$)
 \be
 \omega_{AB}(x) \frac{d x^B}{dt}= - \frac{\partial H}{\partial x^A}\,.
 \ee
This would avoid the need of modifying the covariant spin supplementary condition \eqref{SP=0} by using
instead  a Pryce-Newton-Wigner-type spin-supplementary condition, of the type
 \be
 S_{\mu \nu} (I^\nu + E^\nu_0)=0\,,
 \ee
 as is often used \cite{Barausse:2009aa}. We note in passing that Souriau's result \eqref{sigma9} shows that it is enough to parametrize the spin vector by usual angles $\theta, \phi$ in the rest frame of the particle to obtain for the
 spin-part of $\sigma$ the canonical-like form $  s \sin \theta d\theta \wedge d \phi$, corresponding to the
 standard spin Poisson brackets  $ \{{\hat S}^i, {\hat S}^j \}= \varepsilon_{ijk}  {\hat S}^k$, with $i,j,k=1,2,3$,
 and the constraint $\delta_{ij}{\hat S}^i {\hat S}^j= s^2$.
 [However, this simple property is lost when the mass $m$ becomes spin-dependent, as discussed below.]

This being said, the  need of constructing accurate analytical gravitational-wave templates  for spinning
binary systems (and to be able to compare them to numerical simulations) might be best realized in practice
by working with fully canonical coordinates $q^i, p_i, {\hat S}^i$. Let us emphasize that the accurate description
of the gravitational interaction of spinning compact bodies (neutron stars or black holes) needs to go beyond
the simple pole-dipole approximation described by the Mathisson-Papapetrou equations \eqref{MP} by including
the gravitational couplings of the higher multipole moments. These couplings were formally included in Mathisson's
original work (dealing with a skeletonized worldline), and have been thoroughly investigated (for extended bodies)
by Dixon \cite{Dixon1979}. For instance, if we keep only the effect of the quadrupole moment $J^{\mu\nu\alpha\beta}$,
the equations of motion \eqref{MP} become amplified into
\begin{eqnarray} \label{MPquad}
     \frac{ \nabla S^{\mu \nu }}{ds} &=&P^\mu U^\nu -P^\nu U^\mu+ \frac43 {R^{[\mu}}_{\rho\alpha\beta} J^{\nu] \rho\alpha\beta}\,, \nonumber \\
   \frac{\nabla P^\mu}{ds} &=&- \frac12 {R^\mu}_{\nu \alpha \beta} U^\nu S^{\alpha \beta} -\frac16 \nabla^\mu R_{\nu \rho\alpha\beta} J^{\nu \rho\alpha\beta}\,.  \nonumber \\
\end{eqnarray}
In addition, the quadrupole moment $J^{\mu\nu\alpha\beta}$ (taken here to have the symmetries of the Riemann tensor) is, in general, the sum of two separate contributions: (i) a spin-induced one, symbolically of the form 
$J_{\rm spin} \sim C_2 S^2$, with some body-dependent coefficient $C_2$,
and, (ii), a tidal-induced one, given in the adiabatic approximation by $J_{\rm tidal} \sim \mu_2  { E}$, 
where ${ E}_{\mu\nu}$ denotes the electric
projection of the Riemann tensor (in the rest-frame defined by the relevant spin condition), and where $\mu_2$ is
a (body-dependent) tidal-polarizability coefficient. When using the covariant condition \eqref{SP=0} we have
\be
{ E}_{\mu\nu} \equiv - \frac{P^\alpha P^\beta}{P^2} R_{\alpha \mu \beta \nu}\,.
\ee
For neutron stars, the  quadrupole-level coupling coefficients
$C_2$ and $\mu_2$ are both non-zero and depend on the equation of state. An action describing the
effect of the tidal-induced quadrupole is then $\frac{\mu_2}4 \int ds { E}_{\mu\nu} { E}^{\mu\nu}$.
Note that this action is quadratic in the curvature tensor. By contrast, the spin-induced quadrupolar
coupling can be described by modifying the mass coefficient $m$ entering K\"unzle's prepotential
\eqref{cartan10} (through the relation $P^\mu= m I^\mu$) by a suitable spin-quadrupole modification, 
say\footnote{Note that Ref. \cite{Kuenzle:1972uk} considered such prepotentials involving a field-dependent
mass when discussing spinning particles having a magnetic dipole $\mu$, with ${\mathcal M}(X,S)=m+ \frac12 \mu F_{\mu\nu} S^{\mu\nu}$; the term $\frac12 \mu F_{\mu\nu} S^{\mu\nu}= \mu  {\bf B} \cdot {\bf S}$
being the familiar spin contribution to the rest energy in a magnetic field.}
\be \label{cartan10n}
\varpi_{10}=  {\mathcal M}(X,S) I_\mu  dX^\mu- s e_K  \cdot\nabla e_L\,,
\ee
where ${\mathcal M}(X,S)$ is the (positive) square root of \cite{Porto:2008tb,Vines:2016unv}
\bea \label{M2}
{\mathcal M}^2(X,S)&=& m^2 + \frac{C_2}{4} R_{\mu\nu\alpha\beta} S^{\mu\nu} S^{\alpha\beta} +O(S^3) \nonumber\\
&=& m^2 - C_2 E_{\mu\nu} S^{\mu} S^{\nu} +O(S^3)\,.
\eea
Here, $m$ is a constant mass (which includes the rotational kinetic energy linked to the constant $S^2$),
$C_2$ is a dimensionless number, and $S^\mu= s J^\mu$ is the (Pauli-Lubanski) spin vector.
Note that ${\mathcal M}^2$ (as well as  ${\mathcal M}$ at the $O(S^3) $ accuracy) is linear in the curvature
tensor.
For black-holes, the numerical value of $C_2$ is simply $C_2^{\rm BH}=1$,
while the {\it effacing principle} 
suggests that the tidal-polarizability coefficients of black holes vanish \cite{DamourLH}, so that, in particular 
$\mu_2^{\rm BH}=0$.

In the case of  spinning black holes (described,  when isolated, by the Kerr metric) the generalization of Eq. \eqref{M2}
to the gravitational coupling of the infinite sequence of  (even and odd) spin-induced multipole moments is known
(see, e.g., \cite{Levi:2015msa}). It has been shown in \cite{Vines:2017hyw} that this infinite sequence of couplings
linear in the curvature and its derivatives could be elegantly exponentiated by introducing (in a local frame) the 
matrix operator $(a * \partial)^\mu_{\;\;\nu})={\eta^\mu}_{\nu \alpha\beta} a^\alpha \partial^\beta$ (with $a^\alpha\equiv S^\alpha/m$), which is related to the Janis-Newman complex shift \cite{Newman:1965tw}.

Let us finally mention that the (orbital and spin) dynamics of binary systems of spinning bodies (and
their radiation) has been
recently investigated via the consideration of (classical and/or quantum) {\it scattering observables}. 
For entries in this rapidly growing literature see Refs. \cite{Vines:2017hyw,Bini:2018ywr,Vines:2018gqi,
Guevara:2018wpp,Chung:2018kqs,Guevara:2019fsj,Arkani-Hamed:2019ymq,Bern:2020buy,Kosmopoulos:2021zoq,
Jakobsen:2021zvh,Chen:2021kxt,Cristofoli:2021jas,Aoude:2022trd,Bern:2022kto,Alessio:2022kwv,FebresCordero:2022jts,Cangemi:2022bew,Bjerrum-Bohr:2023jau,Jakobsen:2023hig,Rettegno:2023ghr,Scheopner:2023rzp,Jakobsen:2021lvp,Riva:2022fru,DeAngelis:2023lvf,Brandhuber:2023hhl,Aoude:2023dui}.




\end{document}